\documentclass[twocolumn,english,aps,prl,twocolumn,superscriptaddress,showpacs,floatfix]{revtex4-1}
\usepackage{mathptmx}

\usepackage[T1]{fontenc}
\usepackage[latin9]{inputenc}
\usepackage{textcomp}
\usepackage{amstext}
\usepackage{amssymb}
\usepackage{graphicx}

\makeatletter

\newcommand{\lyxmathsym}[1]{\ifmmode\begingroup\def\b@ld{bold}
  \text{\ifx\math@version\b@ld\bfseries\fi#1}\endgroup\else#1\fi}

\providecommand{\tabularnewline}{\\}

\makeatother

\usepackage{babel}
\begin{document}

\title{Multiple Incommensurate Magnetic States in the Kagome Antiferromagnet
Na$_{2}$Mn$_{3}$Cl$_{8}$}
\author{Joseph A. M. Paddison}
\email{paddisonja@ornl.gov}

\affiliation{Materials Science and Technology Division, Oak Ridge National Laboratory,
Oak Ridge, Tennessee 37831, USA}
\author{Li Yin}
\affiliation{Materials Science and Technology Division, Oak Ridge National Laboratory,
Oak Ridge, Tennessee 37831, USA}
\author{Keith M. Taddei}
\affiliation{Neutron Scattering Division, Oak Ridge National Laboratory, Oak Ridge,
Tennessee 37831, USA}
\author{Malcolm J. Cochran}
\affiliation{Neutron Scattering Division, Oak Ridge National Laboratory, Oak Ridge,
Tennessee 37831, USA}
\author{Stuart A. Calder}
\affiliation{Neutron Scattering Division, Oak Ridge National Laboratory, Oak Ridge,
Tennessee 37831, USA}
\author{David S. Parker}
\affiliation{Materials Science and Technology Division, Oak Ridge National Laboratory,
Oak Ridge, Tennessee 37831, USA}
\author{Andrew F. May}
\email{mayaf@ornl.gov}

\affiliation{Materials Science and Technology Division, Oak Ridge National Laboratory,
Oak Ridge, Tennessee 37831, USA}
\begin{abstract}
The kagome lattice can host exotic magnetic phases arising from frustrated
and competing magnetic interactions. However, relatively few insulating
kagome materials exhibit incommensurate magnetic ordering. Here, we
present a study of the magnetic structures and interactions of antiferromagnetic
Na$_{2}$Mn$_{3}$Cl$_{8}$ with an undistorted Mn$^{2+}$ kagome
network. Using neutron-diffraction and bulk magnetic measurements,
we show that Na$_{2}$Mn$_{3}$Cl$_{8}$ hosts two different incommensurate
magnetic states, which develop at $T_{N1}=1.6$\,K and $T_{N2}=0.6$\,K.
Magnetic Rietveld refinements indicate magnetic propagation vectors
of the form $\mathbf{q}=(q_{x},q_{y},\frac{3}{2})$, and our neutron-diffraction
data can be well described by cycloidal magnetic structures. By optimizing
exchange parameters against magnetic diffuse-scattering data, we show
that the spin Hamiltonian contains ferromagnetic nearest-neighbor
and antiferromagnetic third-neighbor Heisenberg interactions, with
a significant contribution from long-ranged dipolar coupling. This
experimentally-determined interaction model is compared with density-functional-theory
simulations. Using classical Monte Carlo simulations, we show that
these competing interactions explain the experimental observation
of multiple incommensurate magnetic phases and may stabilize multi-\textbf{q
}states. Our results expand the known range of magnetic behavior on
the kagome lattice.
\end{abstract}
\maketitle

\section{Introduction}

Geometrical frustration---the inability of a system to satisfy all
of its pairwise interactions simultaneously---can suppress conventional
magnetic ordering and promote exotic magnetic states \citep{Balents_2010}.
A focus of frustrated-magnetism research has been insulating materials
in which magnetic ions occupy a kagome lattice of corner-sharing triangles,
where strong frustration effects can occur if the interactions are
antiferromagnetic. For example, if antiferromagnetic Heisenberg interactions
couple neighboring spins only, a spin-liquid state is stable down
to extremely low temperatures even in the classical limit \citep{Chalker_1992},
before eventually undergoing octupolar magnetic ordering \citep{Zhitomirsky_2008}.
There is a continuing search for real materials that are candidates
to realize frustrated kagome magnetism \citep{Norman_2016}. In the
quantum ($S=1/2$) limit, notable candidates include herbertsmithite
\citep{Vries_2009,Han_2012} and barlowite \citep{Han_2014,Pasco_2018,Tustain_2018}.
In the classical (large-$S$) limit, probably the most studied candidates
are iron-containing jarosite minerals \citep{Wills_1996,Inami_2000,Nishiyama_2003},
which are often off-stoichiometric \citep{Janas_2020,Bisson_2008}.
Therefore, an important goal is to identify and characterize other
structure types containing kagome lattices, particularly those with
antiferromagnetic interactions, and where the kagome lattice is structurally
undistorted.

Kagome antiferromagnets that exhibit long-range magnetic ordering
may still show strong effects of geometrical frustration. In particular,
the inclusion of further-neighbor interactions can stabilize several
unusual magnetic states instead of conventional collinear antiferromagnetism.
These states include noncollinear $120^{\circ}$ order as well as
many noncoplanar states, which are more stable than collinear antiferromagnets
in large regions of interaction space \citep{Messio_2011}. For certain
exchange interactions, incommensurate magnetic ordering can also be
stabilized; however, the nature of the incommensurate phase is difficult
to determine from simulations \citep{Grison_2020,Li_2022}. Experimental
studies of materials that occupy this part of the interaction space
are therefore important to advance our understanding of kagome magnetism.

\begin{figure}
\includegraphics{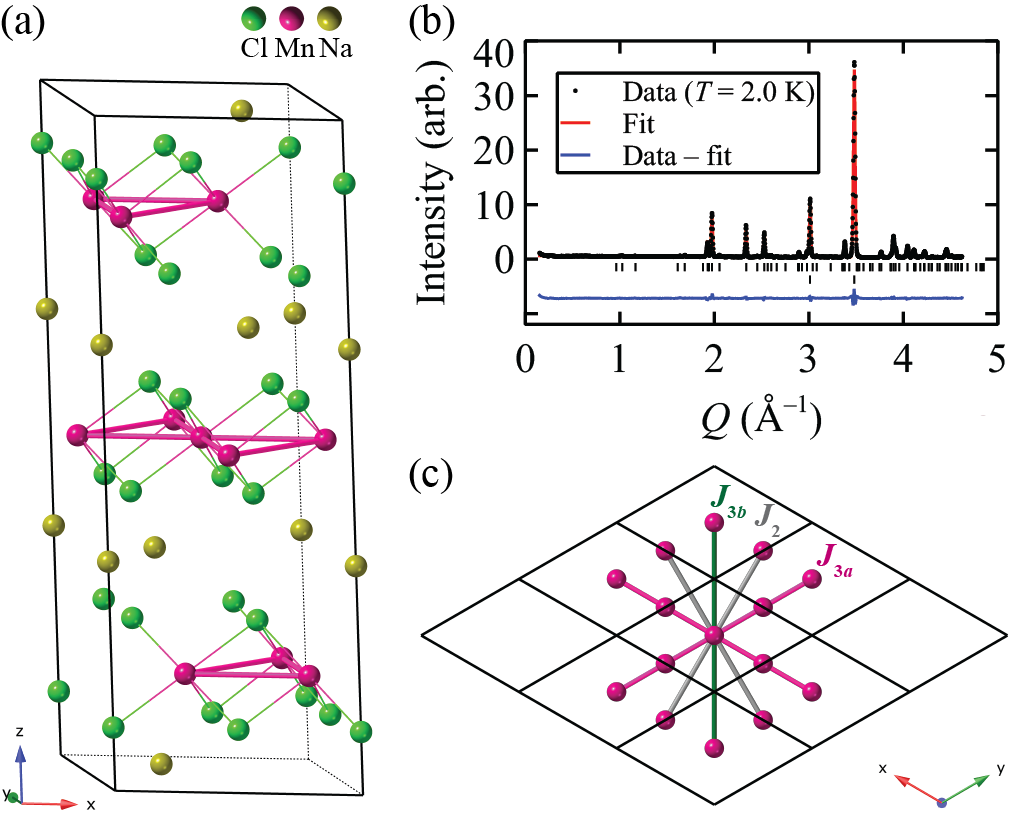}

\caption{\label{fig:crystal_structure}(a) Crystal structure of Na$_{2}$Mn$_{3}$Cl$_{8}$,
showing Mn$^{2+}$ (magenta), Cl$^{-}$(green) and Na$^{+}$(yellow)
ions. (b) Neutron powder diffraction data collected at $T=2$\,K
(black circles), fitted curve from Rietveld refinement (red line),
and data$-$fit (blue line). Experimental data were collected using
the HB-2A diffractometer at ORNL ($\lambda=2.4109\,\text{\AA}$). The
upper and lower tick marks indicate the positions of nuclear Bragg
peaks from Na$_{2}$Mn$_{3}$Cl$_{8}$ and the Cu sample container,
respectively. (c) Magnetic interaction pathways within the kagome
Mn$^{2+}$ layers, showing next-nearest neighbor interactions $J_{2}$
and the two distinct third-neighbor interactions, $J_{3a}$ and $J_{3b}$.
The nearest-neighbor interaction pathway is parallel to $J_{3a}$
at one half of its distance.}
\end{figure}

In this context, we identified Na$_{2}$Mn$_{3}$Cl$_{8}$ as a promising
material for frustrated magnetism on the kagome lattice. This material
was first reported in the 1970s \citep{Loon_1975} and is likely electrically
insulating \citep{Devlin_2022}, but its magnetic structure and interactions
have not previously been studied. Nevertheless, a recent materials
survey highlighted Na$_{2}$Mn$_{3}$Cl$_{8}$ as a candidate frustrated
antiferromagnet \citep{Meschke_2021}. Due to the large magnetic moment
of Mn$^{2+}$with $S=5/2$ and the absence of an orbital contribution
($L=0$), its behavior is expected to be predominantly classical.
The reported crystal structure \citep{Loon_1975} is shown in Figure~\ref{fig:crystal_structure}(a);
its trigonal symmetry (space group $R\bar{3}m$) ensures that the
kagome planes are undistorted. A recent investigation of its bulk
magnetic properties showed multiple magnetic phase transitions below
$2$\,K, and the possibility of a low-temperature structural phase
transition was suggested due to the observation of a broad specific-heat
anomaly around 6\,K \citep{Devlin_2022}. Notably, a structural transition
to a trimerized polar phase is observed in the related $S=1$ kagome
magnet Na$_{2}$Ti$_{3}$Cl$_{8}$ \citep{Hinz_1995,Hanni_2017,Kelly_2019,Paul_2020,Khomskii_2021}.

In this paper, we report magnetic characterization and powder neutron-diffraction
experiments on Na$_{2}$Mn$_{3}$Cl$_{8}$. In agreement with a recent
report \citep{Devlin_2022}, we observe that this material undergoes
two magnetic phase transitions with decreasing temperature. However,
our data do not indicate a measurable crystallographic distortion
at temperatures down to $0.3$\,K, indicating that the undistorted
kagome lattice is preserved. Our powder neutron-diffraction measurements
show that, unusually, the two ordered magnetic states both have incommensurate
magnetic propagation vectors. These data are consistent with single-$\mathbf{q}$
helical magnetic ordering, with an antiferromagnetic stacking of kagome
layers. We show that the development of multiple incommensurate phases
can be explained by a model including Heisenberg exchange interactions
up to third-nearest neighbors and the long-ranged dipolar interaction,
and we estimate the values of the exchange interactions by analyzing
the magnetic diffuse scattering measured above $T_{N1}$. Our results
place Na$_{2}$Mn$_{3}$Cl$_{8}$ in a complex region of the kagome
phase space, in which incommensurate ordering is stabilized by a competition
between short-range ferromagnetic and longer-range antiferromagnetic
interactions. Our interaction model also suggests that Na$_{2}$Mn$_{3}$Cl$_{8}$
deserves further investigation as a potential host of multi-$\mathbf{q}$
spin textures in zero applied magnetic field. 

Our paper is structured as follows. We first introduce the crystal
structure and potential magnetic exchange pathways of Na$_{2}$Mn$_{3}$Cl$_{8}$,
and present thermomagnetic measurements of the bulk magnetic properties.
We then discuss our powder neutron-diffraction data and symmetry-informed
Rietveld analysis, from which the likely single-$\mathbf{q}$ magnetic
structures are determined. Magnetic diffuse-scattering analysis is
employed to parametrize the magnetic interactions that stabilize incommensurate
ordering. We compare and contrast our experimental results with density-functional-theory
calculations. Finally, we discuss the extent to which our experimental
observations can be rationalized using field-theoretical and Monte
Carlo simulations, and conclude by summarizing our results and highlighting
opportunities for future research.

\section{Methods}

\subsection{Sample synthesis}

A polycrystalline sample (mass $2.1$\,g) of Na$_{2}$Mn$_{3}$Cl$_{8}$
was prepared by sealing a stoichiometric mixture MnCl$_{2}$ and NaCl
in SiO$_{2}$ after heating at 250\,$^{\circ}$C under dynamic vacuum
overnight. The mixture was heated to 750\,$^{\circ}$C for several
hours and quenched by removing from the furnace. The sample was ground,
sealed with $\frac{1}{4}$-atm argon, and annealed at $350$\,$^{\circ}$C
for at total of $\approx260$\,h with an additional intermediate
grinding. All handling of this very air-sensitive sample was conducted
in an inert-atmosphere glovebox, and the samples were kept under inert
atmosphere when they were transferred to the vacuum lines for sealing
of the silica tubes.

\subsection{Experimental measurements}

Magnetization measurements were performed using Quantum Design magnetometers
with data below $1.8$\,K collected using a $^{3}$He insert. The
samples were loaded into measurement straws in an inert-atmosphere
glovebox with grease to protect the powders from air during the rapid
loading process.

Neutron-diffraction measurements were performed using the HB-2A powder
diffractometer at the High Flux Isotope Reactor of Oak Ridge National
Laboratory. The incident neutron wavelength $\lambda=2.4109\,\text{\AA}$.
Our powder sample of mass $2.1$\,g was loaded into a 4-mm-diameter
cylindrical Cu container in a He glovebox. The sample was cooled using
a cryostat with a $^{3}$He insert, affording a base temperature of
$\approx0.3$\,K. Counting times were $\approx3$\,hr at 0.3, 0.8,
2.0, 5.0, and 40\,K, and $\approx0.5$\,hr at other temperatures
below $T_{N1}$. The data were corrected for neutron absorption by
the sample \citep{Hewat_1979}. 

\subsection{Magnetic diffuse scattering refinements and field theory\label{subsec:methods_diffuse_field}}

Magnetic diffuse-scattering refinements were performed using the \textsc{Spinteract}
program to refine the values of the exchange interactions \citep{Paddison_2022a}.
The spin Hamiltonian included Heisenberg exchange interactions and
the magnetic dipolar interaction (see Section~\ref{subsec:diffuse}).
The input data were collected at $2$\,K and $5$\,K and were placed
in absolute intensity units (barn\,sr$^{-1}$ Mn$^{-1}$) by normalization
to the nuclear Bragg profile. A high-temperature ($40$\,K) data
set was subtracted from these data. 

The magnetic diffuse scattering $I(Q)$ and bulk susceptibility $\chi T$
were calculated using Onsager reaction-field theory \citep{Paddison_2022a,Logan_1995,Brout_1967},
and a $40$\,K calculation was subtracted from the calculated $I(Q)$.
In this approach, the Fourier transform of the magnetic interactions
is calculated as $J_{ij}^{\alpha\beta}(\mathbf{q})=\sum_{\mathbf{r}}J_{ij}^{\alpha\beta}(\mathbf{r})\exp(-\mathrm{i}\mathbf{q}\cdot\mathbf{r})$,
where $\alpha,\beta$ denote Cartesian spin components, $i,j\in\{1,3\}$
denote sites within the primitive unit cell, and $\mathbf{r}$ is
the vector connecting unit cells containing sites $i$ and $j$. The
interaction matrix formed by the $J_{ij}^{\alpha\beta}(\mathbf{q})$
is diagonalized on a grid of up to $50^{3}$ points in the first Brillouin
zone to determine its eigenvalues $\lambda_{\mu}(\mathbf{q})$ and
eigenvector components $U_{i\mu}^{\alpha}(\mathbf{q})$,
\[
\lambda_{\mu}(\mathbf{q})U_{i\mu}^{\alpha}(\mathbf{q})=\sum_{j}J_{ij}^{\alpha\beta}(\mathbf{q})U_{j\mu}^{\beta}(\mathbf{q}),
\]
where $\mu\in\{1,3\}$ indexes the normal modes. The long-range dipolar
interaction is included using Ewald summation \citep{Enjalran_2004}.
Within a reciprocal-space mean-field approximation, the magnetic propagation
vector of the first ordered state is the wavevector at which $\lambda_{\mu}$
reaches its maximal value. The $I(Q)$ and $\chi T$ are given in
terms of the $\lambda_{\mu}(\mathbf{q})$ and $U_{i\mu}^{\alpha}(\mathbf{q})$,
as described in Ref.~\citep{Paddison_2022a}.

During the refinements, we minimized the function
\begin{equation}
\chi^{2}=\sum_{i}\left(\frac{I_{\mathrm{expt}}^{i}-sI_{\mathrm{calc}}^{i}}{\sigma_{i}}^{2}\right),\label{eq:chi_Sq-1}
\end{equation}
where subscript ``expt'' and ``calc'' indicate measured and calculated
diffuse scattering patterns, respectively, $\sigma$ is an experimental
uncertainty, and $s$ is a refined overall scale factor common to
the neutron-scattering data and the magnetic susceptibility $\chi T$.
The minimization was performed using the \textsc{Minuit }program \citep{James_1975,James_1994}.
To identify local minima in $\chi^{2}$, we performed $25$ refinements
for each model, with different randomly-chosen initial parameter values
in each case.

\subsection{Rietveld refinements}

Rietveld refinements were performed using the \textsc{Fullprof} software
\citep{Rodriguez-Carvajal_1993,Rodriguez-Carvajal_1993a}. A crystal-structure
refinement was first performed at $T=2$\,K ($>T_{N1}$). In addition
to the crystallographic parameters given in Table~\ref{tab:crystallographic},
we refined the intensity scale factor, $2\theta$ zero-offset, peak-shape,
and background parameters. The peak shape was modeled using a pseudo-Voigt
function initialized with the instrument resolution parameters, with
$U$, $V$, and $W$ parameters subsequently refined. The background
was fitted using Chebychev polynomials.

Magnetic Rietveld refinements were performed against $0.3$\,K and
$0.8$\,K data from which the $2$\,K data had been subtracted.
This subtraction isolates the magnetic Bragg signal by subtracting
the nuclear and background contributions, which are essentially unchanged
between $0.3$ and $2$\,K. In the magnetic refinements, asymmetry,
Chebychev background, and magnetic-structure parameters were refined,
as described in Section~\ref{subsec:bragg}; all other parameters
were fixed at the values obtained from the $2$\,K refinement. Magnetic-structure
figures were prepared using the \textsc{Vesta} program \citep{Momma_2008}.

\subsection{Density-functional-theory calculations}

Density functional theory calculations were performed using the all-electron-density
functional code \textsc{Wien2K} \citep{Sjostedt_2000,Blaha_2001}.
The linearized augmented plane wave method \citep{Singh_2006} and
the generalized-gradient approximation of Perdew, Burke, and Ernzerhof
\citep{Perdew_1996} were utilized. The $RK_{\mathrm{max}}$ generated
by the smallest linearized augmented plane wave sphere radius ($R$)
and the interstitial plane-wave cutoff ($K_{\mathrm{max}}$) was set
as 7.0 for good convergence. The muffin-tin radii of Na, Cl, and Mn
atoms were $2.47$\,a.u., $2.14$\,a.u., and $2.49$\,a.u., respectively.
The number of $\mathbf{q}$-points in the full Brillouin zone was
$200$. Lattice parameters of Na$_{2}$Mn$_{3}$Cl$_{8}$ were fixed
to the experimental values of $a=b=7.423$\,$\text{\AA}$ and $c=19.497$\,$\text{\AA}$.
Then, the internal atomic coordinates were relaxed until forces on
all of the atoms were less than $1$\,mRy/bohr, with non-magnetic,
ferromagnetic, and interlayer antiferromagnetic states. It turns out
the relaxed crystal structure with a ferromagnetic state is highly
similar to the experimental crystal structure. However, in the non-magnetic
state, the atomic coordination changes significantly, as Cl atoms
moves towards the Mn layers. The relaxed antiferromagnetic crystal
structure is same as the ferromagnetic crystal structure, but exhibits
lower energy and smaller forces. The energy difference between ferromagnetic
and antiferromagnetic states is $1.19$\,meV/f.u.. We therefore used
the relaxed antiferromagnetic structure to calculate the intralayer
and interlayer magnetic couplings.

\section{Experimental Results}

\subsection{Crystal structure refinement}

\begin{table}
\centering{}%
\begin{tabular}{ccc}
\hline 
\multicolumn{3}{c}{Na$_{2}$Mn$_{3}$Cl$_{8}$, $T=2$\,K}\tabularnewline
\multicolumn{3}{c}{$R\overline{3}m$, $a=7.4249(1)\,\text{\AA},\,c=19.4971(4)\thinspace\lyxmathsym{\AA}$}\tabularnewline
\multicolumn{3}{c}{$B_{\mathrm{overall}}=0.31(6)\,\text{\AA}^{2}$}\tabularnewline
\hline 
Site & Wyckoff & $(x,y,z)$\tabularnewline
\hline 
Na & $6c$ & $(0,0,0.3395(7))$\tabularnewline
Mn & $3b$ & $(0,0,\frac{1}{2}$)\tabularnewline
Cl1 & $6c$ & $(0,0,0.9062(4))$\tabularnewline
Cl2 & $18h$ & $(0.5081(3),0.4919(3),0.0931(2)$)\tabularnewline
\hline 
\end{tabular}\caption{\label{tab:crystallographic}Refined crystallographic parameters of
Na$_{2}$Mn$_{3}$Cl$_{8}$ at $T=2$\,K, obtained from Rietveld
refinement to powder neutron-diffraction data ($\lambda=2.4109\,\text{\AA}$).}
\end{table}

The crystal structure of Na$_{2}$Mn$_{3}$Cl$_{8}$ is shown in Figure~\ref{fig:crystal_structure}(a),
and comprises of triangular Na$^{+}$ layers separating kagome Mn$^{2+}$
layers \citep{Loon_1975,Devlin_2022}. We performed Rietveld refinements
against our $2$\,K and $40$\,K neutron-diffraction data to investigate
the possibility of a crystallographic distortion from the published
structure (space group $R\bar{3}m$). Good agreement was obtained
with the published structural model \citep{Loon_1975} at both temperatures,
except for two very weak peaks at $1.64$ and $2.16\,\text{\AA}^{-1}$
that were not accounted for, and were unchanged between $0.3$ and
$40$\,K. Since these peaks could not be explained by simple multiples
of the crystallographic unit cell, or by possible impurity phases
(NaCl, MnCl$_{2}$, or NaMnCl$_{3}$), we concluded that the sample
or its environment contained a small fraction of unknown impurity.
Our results do not show evidence for any structural phase transition
between $2$\,K and $40$\,K, indicating that the broad $\sim$$6$\,K
specific-heat anomaly reported previously \citep{Devlin_2022} is probably due to magnetic
ordering of a minor NaMnCl$_{3}$ impurity phase (a possibility noted
in Ref.~\citep{Devlin_2022}).

Each nearest-neighbor Mn--Mn bond is bridged by a Cl1 ion and a Cl2
ion, which provide the nearest-neighbor superexchange pathways. The
Mn--Cl1--Mn and Mn--Cl2--Mn bond angles are $92.42^{\circ}$ and
$94.03^{\circ}$, respectively. Since these values are close to $90^{\circ}$,
the Goodenough-Kanamori rules predict weak ferromagnetic nearest-neighbor
exchange interactions. Further-neighbor interactions have more complicated
pathways and, consequently, are difficult to predict. In particular, there are two inequivalent third-neighbor exchange pathways with the same interatomic separation [Figure~\ref{fig:crystal_structure}(c)].

\subsection{Thermomagnetic measurements}

\begin{figure}
\includegraphics[width=8.6cm]{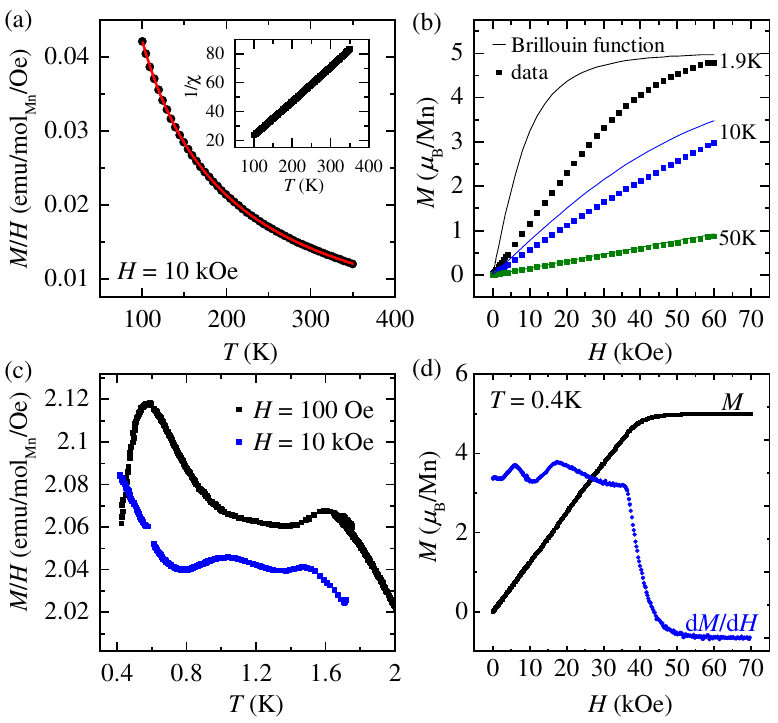}
\centering{}\caption{\label{fig:bulk_magnetism}Overview of bulk magnetic measurements.
(a) High-temperature powder magnetic susceptibility $\chi\approx M/H$
measured in applied field of $10$\,kOe (black squares), and Curie-Weiss
fit (red line). (b) Low-temperature magnetic susceptibility measured
in applied fields of $100$\,Oe (black points) and $10$\,kOe (blue
points), indicating magnetic phase transitions at approximately $0.6$
and $1.6$\,K. (c) Dependence of magnetization $M$ on applied field
$H$ at temperatures of $1.9$, $10$, and $50$\,K (black, blue,
and green squares, respectively). (d) Dependence of magnetization
$M$ on applied field $H$ at $0.4$\,K (black squares) and its field
derivative (blue circles).}
\end{figure}

Our high-temperature bulk magnetic susceptibility measurements and
Curie-Weiss fits are shown in Figure~\ref{fig:bulk_magnetism}(a).
They reveal an effective magnetic moment of $5.99$\,$\mu_{\mathrm{B}}$,
close to the spin-only value of $5.92$\,$\mu_{\mathrm{B}}$ for
Mn$^{2+}$, and a Weiss temperature of $\theta_{\mathrm{W}}=-4.6(1)$\,K,
indicating net antiferromagnetic interactions. An anomaly is observed
at $T_{N1}\approx1.6$\,K in our low-temperature magnetic susceptibility
data, which is suppressed to lower temperature with increasing applied
field, consistent with a magnetic ordering transition {[}Figure~\ref{fig:bulk_magnetism}(b){]}.
The \textquotedblleft frustration parameter\textquotedblright , $f=\theta_{\mathrm{W}}/T_{N1}\approx3$,
indicates a relatively small degree of frustration. Since the nearest-neighbor
kagome antiferromagnet is highly frustrated, this result hints at
the presence of significant further-neighbor couplings or anisotropies;
however, the nature of these couplings cannot be determined from bulk
characterization data alone. Interestingly, and consistent with Ref.~\citep{Devlin_2022},
we also observe a second magnetic-susceptibility anomaly at $T_{N2}\approx0.6$\,K
{[}Fig.~\ref{fig:bulk_magnetism}(b){]}, suggesting a multi-stage
magnetic ordering process. Such behavior is unusual and hints that,
despite the relatively small value of $f$, the frustrated topology
of the kagome lattice may cause several magnetic structures to be
nearly degenerate. Figure.~\ref{fig:bulk_magnetism}(c) shows the
low temperature field dependence of the magnetization, which does
not follow the Brillouin function, in qualitative agreement for theoretical
predictions for the kagome lattice \citep{Nakano_2015}. Figure~\ref{fig:crystal_structure}(a)
shows the field derivative of the magnetization, $dM/dH$. Several
anomalies are observed in $dM/dH$ below $T_{N2}$ at small applied
fields, suggesting that the magnetic ground state is fragile to external
perturbations.

\subsection{Overview of neutron data}

\begin{figure}
\includegraphics{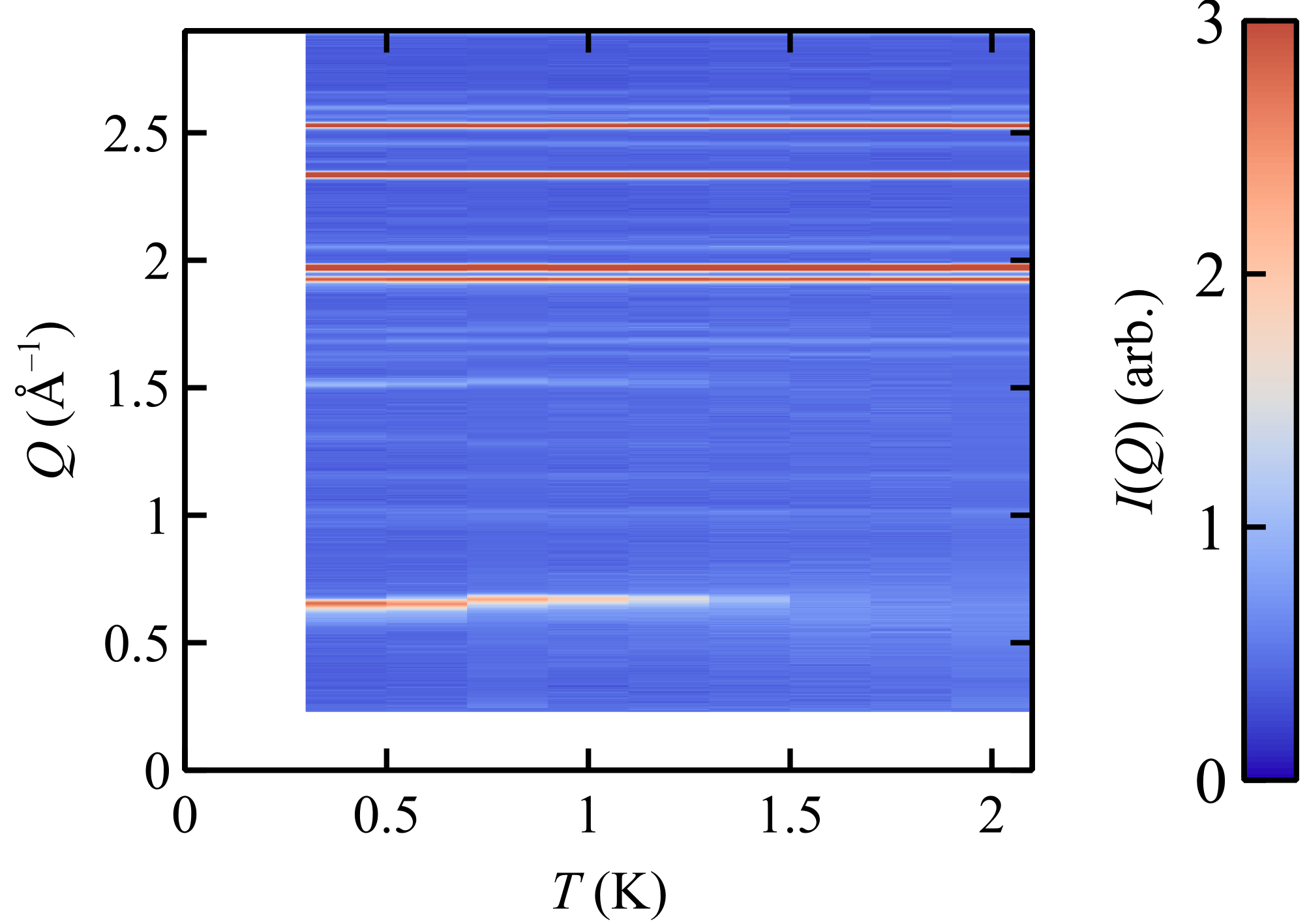}
\centering{}\caption{\label{fig:neutron_overview}Overview of neutron-diffraction data,
showing diffraction intensity in false color as a function of temperature
$T$ and wavevector $Q$. Magnetic ordering at $T_{N1}\approx1.6$\,K
is indicated by the appearance of new (magnetic) Bragg peaks, and
the magnetic phase transition at $T_{N2}\approx0.6$\,K is indicated
by a change in position of these peaks. Nuclear peaks, such as the
four intense peaks at $Q>1.8$\,$\text{\protect\AA}^{-1}$, do not
change position with temperature.}
\end{figure}

We performed neutron-diffraction measurements to obtain microscopic
insight into the magnetic interactions and structures of Na$_{2}$Mn$_{3}$Cl$_{8}$
(see Methods). An overview of the temperature dependence of our neutron
data is shown in Figure~\ref{fig:neutron_overview}(a). Several new
Bragg peaks appear below $T_{N1}\approx1.6$\,K, most prominently
at wavevectors of approximately $0.6$ and $1.5\,\text{\AA}^{-1}$.
We identify these as magnetic Bragg peaks arising from the onset of
long-range magnetic ordering, since they appear at the same temperature
as the magnetic-susceptibility anomaly at $T_{N1}$. Interestingly,
the positions of the magnetic Bragg peaks suddenly shift at $T_{N2}\approx0.6$\,K,
revealing that the second phase transition involves a change in magnetic
propagation vector. At temperatures above $T_{N1}$, broad magnetic
diffuse scattering features can be seen, indicating the development
of short-range magnetic correlations as $T_{N1}$ is approached from
above. We discuss the Bragg and diffuse magnetic scattering in Section~\ref{subsec:bragg}
and \ref{subsec:diffuse}, respectively.

\subsection{Magnetic structures from Rietveld refinements\label{subsec:bragg}}

We first discuss possible ordered magnetic structures of Na$_{2}$Mn$_{3}$Cl$_{8}$,
as determined by analyzing the magnetic Bragg profiles obtained at
temperatures below $T_{N1}$.

We used the program \textsc{KSearch} of the \textsc{Fullprof} suite
\citep{Rodriguez-Carvajal_1993,Rodriguez-Carvajal_1993a} to identify
possible propagation vectors at $0.3$\,K ($T<T_{N2}$) and $0.8$\,K
($T_{N2}<T<T_{N1}$). The positions of 10 magnetic peaks (at $0.8$\,K)
and 15 magnetic peaks (at $0.3$\,K) were provided as input, and
a systematic search of candidate propagation vectors $\mathbf{q}=q_{x}\text{\textbf{a}}^{\ast}+q_{y}\text{\textbf{b}}^{\ast}+q_{z}\text{\textbf{c}}^{\ast}$
was performed, starting with those that lie on a symmetry point, line,
or plane of the Brillouin zone. However, none of the high-symmetry
propagation vectors was compatible with the observed Bragg positions,
at either temperature. The best-fit propagation vectors were instead
of the form $(q+\delta,q-\delta,\frac{3}{2})$, with $\delta\ll q$.
We obtain $(q,\delta)_{\mathrm{}}\approx(0.29,0.02)$ at $0.8$\,K,
and $(q,\delta)\approx(0.27,0.06)$ at $0.3$\,K; precise values
are given in Table~\ref{tab:magnetic_refinements}. These propagation
vectors lie on a general position, but they are close to the high-symmetry
$(q,q,\frac{3}{2})$ plane.

Having determined possible propagation vectors, we used the program
\textsc{Sarah }\citep{Wills_2000} to identify symmetry-allowed magnetic
structures. The primitive unit cell contains three Mn$^{2+}$ sites,
with fractional coordinates $\text{\textbf{r}}_{1}=(\frac{1}{2},0,\frac{1}{2})$,
$\text{\textbf{r}}_{2}=(0,\frac{1}{2},\frac{1}{2})$, and $\text{\textbf{r}}_{3}=(\frac{1}{2},\frac{1}{2},\frac{1}{2})$
with respect to the conventional axes $\mathbf{a}$, $\mathbf{b}$,
$\mathbf{c}$. Each site has three magnetic degrees of freedom, which
are not further constrained by symmetry. We choose these as basis-vector
components along orthonormal axes $\mathbf{q}_{\parallel}$, $\mathbf{q}_{\perp}$,
$\hat{\mathbf{c}}$, where $\hat{\mathbf{c}}$ is parallel to the
$c$-axis, $\mathbf{q}_{\parallel}$ is parallel to the projection
of $\mathbf{q}$ in the $ab$-plane, and $\mathbf{q}_{\perp}=\hat{\mathbf{c}}\times\mathbf{q}_{\parallel}$
is perpendicular to $\mathbf{q}_{\parallel}$ and $\text{\textbf{c}}$.
These structures are amplitude-modulated spin-density waves (sine
structures), with different spin magnitudes and orientations for each
site,
\begin{equation}
\text{\ensuremath{\mu}}_{\textrm{sine}}^{j}(\mathbf{R})\propto\left(\mu_{\mathbf{q}_{\parallel}}^{j},\mu_{\mathbf{q}_{\perp}}^{j},\mu_{\mathbf{c}}^{j}\right)\exp(-2\pi\mathrm{i}\mathbf{q}\cdot\mathbf{R})+\mathrm{c.c.},\label{eq:sine}
\end{equation}
where $\mathbf{q}$ denotes the propagation vector, $\mathbf{R}$
denotes a lattice vector, $j\in\{1,3\}$ labels sites within the unit
cell, and $\mu_{\mathbf{q}_{\parallel}},\mu_{\mathbf{q}_{\perp}},\mu_{\mathbf{c}}$
are basis-vector components. Alternatively, it is possible to construct
helical structures such as
\begin{equation}
\text{\ensuremath{\mu}}_{\mathrm{helix}}^{j}(\mathbf{R})\propto\left(\mu_{\mathbf{q}_{\parallel}}^{j},\mathrm{i}\mu_{\mathbf{q}_{\perp}}^{j},0\right)\exp(-2\pi\mathrm{i}\mathbf{q}\cdot\mathbf{R})+\mathrm{c.c.},\label{eq:helix}
\end{equation}
where, in this case, the spin plane is perpendicular to the $c$-axis.
The ordered magnetic-moment length can be identical on all sites in
the crystal in a helical structure, for example if $\mu_{\mathbf{q}_{\parallel}}=\mu_{\mathbf{q}_{\perp}}$
in Eq.~(\ref{eq:helix}).

Due to the relatively large number of variable parameters and the
limitations of powder data, we make two assumptions when testing candidate
magnetic structures. First, we only consider structures that order
with a single propagation vector (single-$\mathbf{q}$ structures).
While multi-$\mathbf{q}$ structures are possible, they cannot generally
be distinguished from single-$\mathbf{q}$ structures by powder diffraction
\cite{Kouvel_1963}. Second, we initially assume that the basis vectors at sites $\mathbf{r}_{1}$,
$\mathbf{r}_{2}$ and $\mathbf{r}_{3}$ are parallel; this assumption
is reasonable because the interactions between nearest and next-nearest
neighbors are ferromagnetic, as we will show in Section~\ref{subsec:diffuse}.
Magnetic-structure models were tested against the magnetic Bragg profile
using Rietveld refinement (see Methods).

\begin{figure*}
\begin{centering}
\includegraphics[scale=0.83]{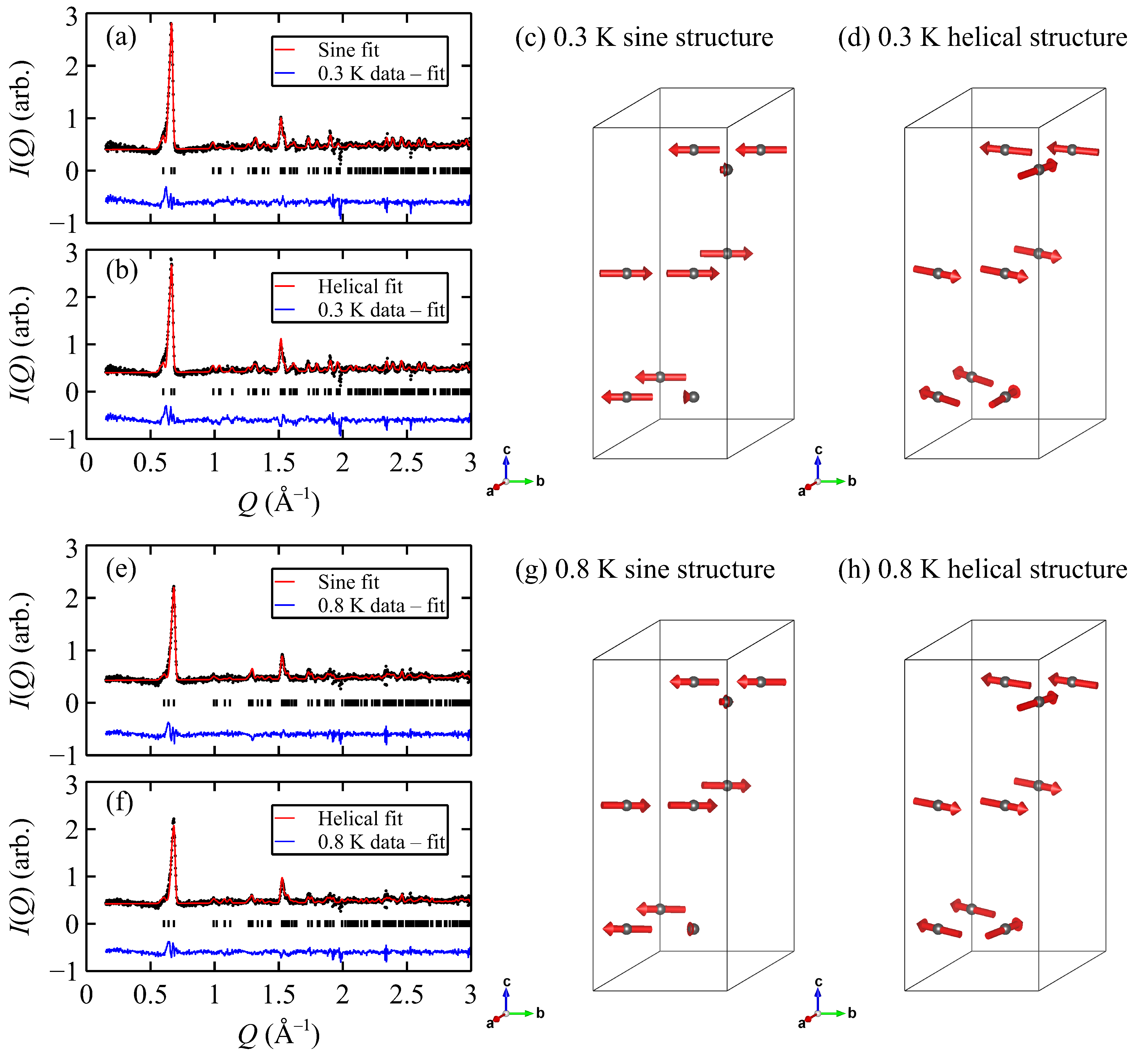}
\par\end{centering}
\centering{}\caption{\label{fig:neutron_rietveld}(a,b) Magnetic neutron-diffraction data
at $T=0.3$\,K with $2$\,K data subtracted (black circles), Rietveld
fits for the sine (a) and helical (b) structures (red lines), and
data$-$fit curves (blue lines). (c,d) Possible single-$\mathbf{q}$
magnetic structures at $T=0.3$\,K, showing sine (c) and helical
(d) candidates. (e,f) Magnetic neutron diffraction data at $T=0.8$\,K
with $2$\,K data subtracted (black circles), Rietveld fits for the
sine (e) and helical (f) structures (red lines), and data$-$fit curves
(blue lines). (g,h) Possible single-$\mathbf{q}$ magnetic structures
at $T=0.8$\,K, showing sine (g) and helical (h) candidates.}
\end{figure*}

\begin{table}
\begin{centering}
\begin{tabular}{c|c|c|c|c|c}
\hline 
\multicolumn{6}{c}{Na$_{2}$Mn$_{3}$Cl$_{8}$, magnetic}\tabularnewline
\multicolumn{6}{c}{$\mathbf{q}=(0.3282(3),0.2117(3),\frac{3}{2})$ at $T=0.3$\,K}\tabularnewline
\multicolumn{6}{c}{$\mathbf{q}=(0.3102(4),0.2646(4),\frac{3}{2})$ at $T=0.8$\,K}\tabularnewline
\hline 
\multicolumn{1}{c}{} & \multicolumn{1}{c}{} & \multicolumn{1}{c}{} & \multicolumn{1}{c}{} & \multicolumn{1}{c}{} & \tabularnewline
\hline 
$T$ (K) & Structure & $\mu_{\mathbf{q}\parallel}$\,($\mu_{\mathrm{B}}$) & $\mu_{\mathbf{q}\perp}$\,($\mu_{\mathrm{B}}$) & $\mu_{\mathbf{c}}$\,($\mu_{\mathrm{B}}$) & $R_{\mathrm{wp}}$ (\%)\tabularnewline
\hline 
$0.3$ & sine & $1.15(16)$ & $6.02(4)$ & $-0.48(22)$ & $25.7$\tabularnewline
$0.8$ & sine & $0.25(18)$ & $5.32(4)$ & $-0.86(21)$ & $30.5$\tabularnewline
\hline 
\multicolumn{1}{c}{} & \multicolumn{1}{c}{} & \multicolumn{1}{c}{} & \multicolumn{1}{c}{} & \multicolumn{1}{c}{} & \tabularnewline
\hline 
$T$ (K) & Structure & $\mu_{\mathrm{ord}}$\,($\mu_{\mathrm{B}}$) & $\Delta\phi$\,($\lyxmathsym{\textdegree}$) & $\theta$\,($\lyxmathsym{\textdegree}$) & $R_{\mathrm{wp}}$ (\%)\tabularnewline
\hline 
$0.3$ & $\mathbf{q}_{\parallel}\mathbf{c}$-helix & $5.35(6)$ & $0^{\ast}$ & $0^{\ast}$ & $37.9$\tabularnewline
 & $\mathbf{q}_{\perp}\mathbf{c}$-helix & $4.85(5)$ & $0^{\ast}$ & $0^{\ast}$ & $29.3$\tabularnewline
 & $\mathbf{ab}$-helix & $4.77(5)$ & $0^{\ast}$ & $0^{\ast}$ & $28.2$\tabularnewline
 & $\mathbf{ab}$-helix & $4.55(5)$ & $22(2)$ & $0^{\ast}$ & $27.9$\tabularnewline
 & helix & $4.84(6)$ & $0^{\ast}$ & $\approx35$ & $27.5$\tabularnewline
$0.8$ & $\mathbf{q}_{\parallel}\mathbf{c}$-helix & $4.66(6)$ & $0^{\ast}$ & $0^{\ast}$ & $41.5$\tabularnewline
 & $\mathbf{q}_{\perp}\mathbf{c}$-helix & $4.20(5)$ & $0^{\ast}$ & $0^{\ast}$ & $33.0$\tabularnewline
 & $\mathbf{ab}$-helix & $4.23(5)$ & $0^{\ast}$ & $0^{\ast}$ & $31.9$\tabularnewline
 & $\mathbf{ab}$-helix & $4.06(5)$ & $91(3)$ & $0^{\ast}$ & $30.0$\tabularnewline
 & helix & $4.23(5)$ & $0^{\ast}$ & $\lesssim30$ & $31.9$\tabularnewline
\hline 
\end{tabular}
\par\end{centering}
\caption{\label{tab:magnetic_refinements}Refined values of magnetic-structure
parameters for different single-$\mathbf{q}$ models, and corresponding
goodness-of-fit metric $R_{\mathrm{wp}}$. The refined parameters
are defined in the text.}
\end{table}

We first considered amplitude-modulated sine structures. The assumption
of parallel basis vectors reduces the number of refined parameters
from 9 to 3. Sine structures yield excellent agreement with our data
at both 0.3 and 0.8\,K, as shown in Figure~\ref{fig:neutron_rietveld}(a)
and (e), respectively. The magnetic moment is predominantly oriented
along $\mathbf{q}_{\perp}$ for the corresponding structures, which
are shown in Figure~\ref{fig:neutron_rietveld}(c) and (g), respectively.
The refined parameter values and goodness-of-fit metric $R_{\mathrm{wp}}$
are given in Table~\ref{tab:magnetic_refinements}. To determine
if sine structures are physically reasonable, we calculated the maximum
value of the ordered magnetic moment, $\mathrm{max}(\mu_{\mathrm{ord}})$.
For a spin-only ion, this value should not normally exceed $2S\thinspace\mu_{\mathrm{B}}$
($=5.0\thinspace\mu_{\mathrm{B}}$ for Mn$^{2+}$). This expectation
is confirmed by the low-temperature magnetization of Na$_{2}$Mn$_{3}$Cl$_{8}$,
which saturates to approximately $5\thinspace\mu_{\mathrm{B}}$ per
Mn$^{2+}$ {[}Figure~\ref{fig:bulk_magnetism}(d){]}. Unfortunately,
we find $\mathrm{max}(\mu_{\mathrm{ord}})\gg5.0\mu_{\mathrm{B}}$
for the refined sine structures: $\mathrm{max}(\mu_{\mathrm{ord}})=6.15(7)\thinspace\mu_{\mathrm{B}}$
at $0.3$\,K, and $5.40(7)\thinspace\mu_{\mathrm{B}}$ at $0.8$\,K.
These values are physically unreasonable, suggesting that the correct
structures of Na$_{2}$Mn$_{3}$Cl$_{8}$ are not single-$\mathbf{q}$
sine structures.

Circular helices are promising alternative structures, since all sites
have equal magnetic moment lengths. Initially, we consider circular
helices with magnetic moments in either the $\mathbf{q}_{\parallel}\mathbf{c}$
plane, the $\mathbf{q}_{\perp}\mathbf{c}$ plane, or the $\mathbf{ab}$
plane (equivalent to the $\mathbf{q}_{\perp}\mathbf{q}_{\parallel}$
plane). At both $0.3$ and $0.8$\,K, the best fit is obtained for
the $\mathbf{ab}$-helix, with slightly worse agreement for the $\mathbf{q}_{\perp}\mathbf{c}$-helix
{[}Table~\ref{tab:magnetic_refinements}{]}. The $\mathbf{q}_{\parallel}\mathbf{c}$-helix
yields much worse agreement than the other structures, so we do not
consider it further. The fits for $\mathbf{ab}$-helices at $0.3$
and $0.8$\,K are shown in Figure~\ref{fig:neutron_rietveld}(b)
and (f), respectively, and the corresponding structures are shown
in Figure~\ref{fig:neutron_rietveld}(d) and (h). The agreement with
the data is very good, although marginally worse than for the corresponding
sine structures. Importantly, however, the refined values of $\mu_{\mathrm{ord}}$
are now physically reasonable, with a maximum value of $4.77(5)\thinspace\mu_{\mathrm{B}}$
at $0.3$~K. This result favors the helical structures.

We tested two variations of the helical structures in an effort to
improve the fit quality. First, we considered the $\mathbf{ab}$-helix
and relaxed our previous assumption of parallel basis vectors, by
refining a clockwise rotation $\Delta\phi$ of the basis vector at
position $\mathbf{r}_{3}=(\frac{1}{2},\frac{1}{2},\frac{1}{2})$ about
the $c$-axis. The optimal fit is obtained for relatively small $\Delta\phi\approx20^{\circ}$
at $0.3$ K, and substantial $\Delta\phi\approx90^{\circ}$ at $0.8$
K. Second, we maintain the assumption that the basis vectors are parallel,
but vary the spin plane as $\mathbf{q}_{\perp}(\mathbf{q}_{\parallel}\cos\theta+\mathbf{c}\sin\theta)$.
At $0.3$\,K, a minimum in $R_{\mathrm{wp}}$ occurs for $\theta\approx35^{\circ}$,
whereas at $0.8$\,K, fit quality is essentially unchanged for all
$\theta\lesssim30^{\circ}$. Each of these variations yields a similar
or slightly improved fit compared to the simple $\mathbf{ab}$-helix
(see Table \ref{tab:magnetic_refinements}).

\begin{figure}
\includegraphics{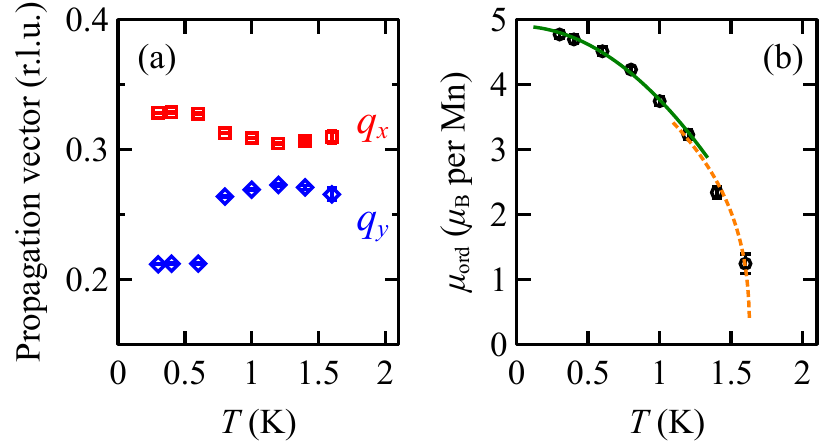}

\caption{\label{fig:tdep}Temperature dependence of refined parameters for
the $\mathbf{ab}$-helix structure with collinear basis vectors. (a)
Temperature evolution of the magnetic propagation vector $(q_{x},q_{y},\frac{3}{2})$,
showing $q_{x}$ (red squares) and $q_{y}$ (blue diamonds). (b) Temperature
evolution of the ordered magnetic moment $\mu_{\mathrm{ord}}$ (black
circles). The solid green line is a fit to $\mu_{\mathrm{sat}}(1-cT^{2})$,
where $\mu_{\mathrm{sat}}=4.90(4)\,\mu_{\mathrm{B}}$ and $c=0.23(1)$\,K$^{-2}$.
The dotted orange line is a fit to the critical form for a three-dimensional
Heisenberg magnet, $m(T_{N1}-T)^{0.365}$, where $T_{N1}=1.63(1)$\,K
and $m=4.3(1)\,\mu_{\mathrm{B}}$.}
\end{figure}

Figure~\ref{fig:tdep} shows the temperature evolution of the refined
parameter values for the $\mathbf{ab}$-helix with collinear basis
vectors. A discontinuity in the propagation vector is apparent at
$T_{N2}$ {[}Figure~\ref{fig:tdep}(a){]}. No such anomaly is apparent
in the refined value of the ordered magnetic moment, which increases
smoothly on cooling the sample below $T_{N1}$ {[}Figure~\ref{fig:tdep}(b){]}.
The temperature dependence of this order parameter at low temperatures
($T\leq1.2$\,K) is consistent with the phenomenological form $\mu_{\mathrm{ord}}\propto1-cT^{2}$
for a three-dimensional magnet with half-integer spin \cite{Kobler_2003}. Its temperature
dependence for $1.2\leq T\leq1.6$\,K is consistent with critical
form for a three-dimensional Heisenberg magnet, $\mu_{\mathrm{ord}}\propto(T_{N1}-T)^{0.365}$,
although the small number of data points precludes fitting the critical
exponent.

In conclusion, our powder-diffraction data are well explained by circular
helical magnetic structures. Basis vectors are close to the $\mathbf{ab}$
plane and nearly collinear at $0.3$\,K, with a possibility of greater
noncollinearity at $0.8$\,K. We emphasize, however, that the possibility
of multi-$\mathbf{q}$ structures cannot be ruled out, and we discuss
this further in Section~\ref{subsec:incommensurate}.

\subsection{Magnetic interactions from diffuse scattering\label{subsec:diffuse}}

\begin{figure*}
\begin{centering}
\includegraphics[scale=0.83]{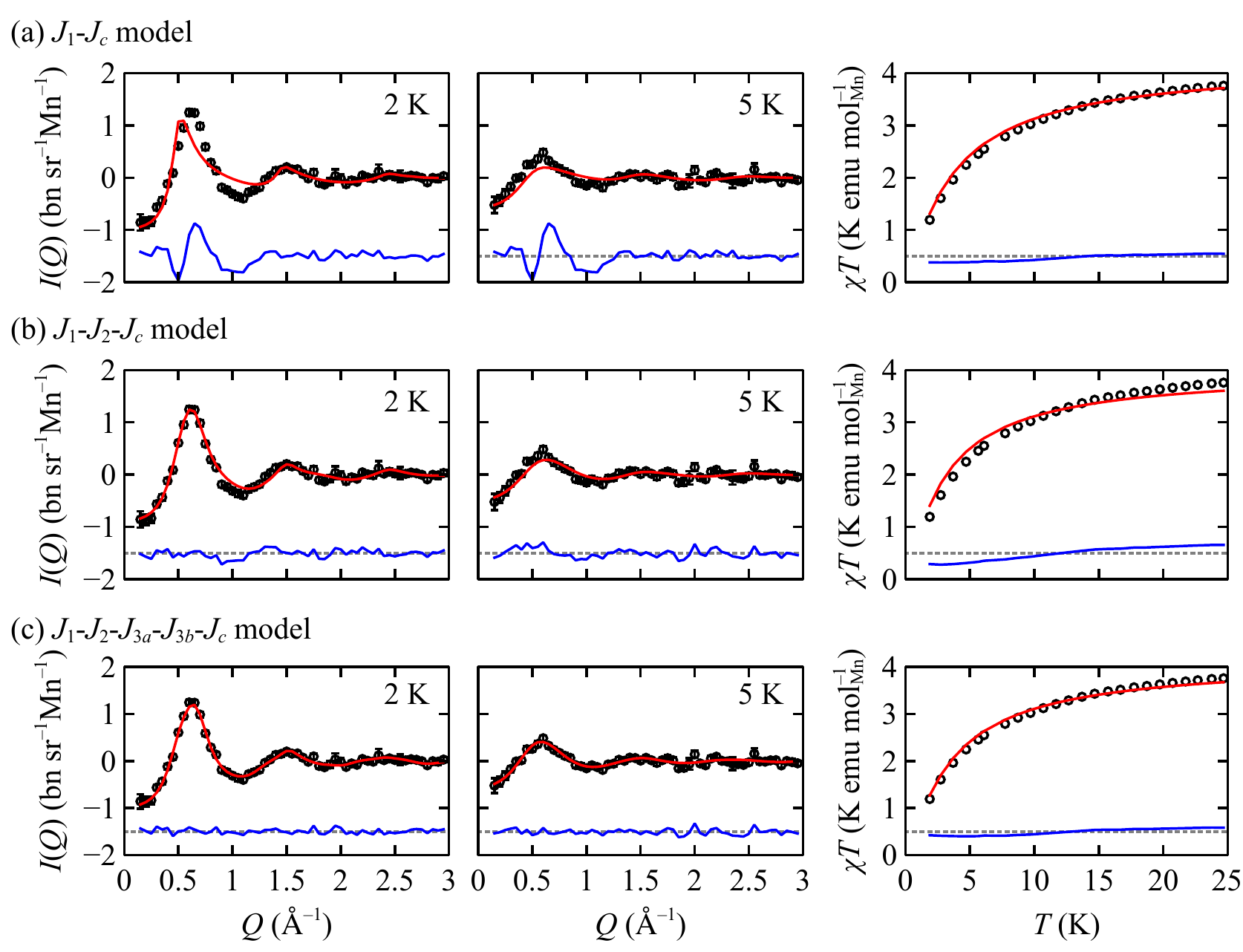}
\par\end{centering}
\centering{}\caption{\label{fig:diffuse}Experimental data (black circles), model fits
(red lines), and difference (data$-$fit) curves (blue lines) at temperatures
above $T_{N1}$. Left and center columns show neutron-scattering data
collected at 2\,K and $5$\,K, respectively, and the right column
shows magnetic susceptibility ($\chi T$) data. A high-temperature
($40$\,K) data set has been subtracted from the neutron data shown.
The models (a), (b), and (c) are described in the main text, and the
parameter values for each fit are given in Table\,\ref{tab:diffuse}.}
\end{figure*}

We seek to parametrize the spin Hamiltonian of Na$_{2}$Mn$_{3}$Cl$_{8}$
by analyzing the diffuse magnetic scattering measured above $T_{N1}$.
This approach is an alternative to spin-wave analysis of inelastic
neutron-scattering data, and has recently been applied to several
frustrated antiferromagnets \citep{Samarakoon_2020,Scheie_2021,Welch_2022}.
The magnetic diffuse scattering measured at $2$\,K and $5$\,K (with $40$\,K data subtracted)
is shown in Figure~\ref{fig:diffuse}. Diffuse magnetic peaks are
sharper at $2$\,K than at $5$\,K, consistent with an increase
in the magnetic correlation length on cooling the sample. The bulk
magnetic susceptibility expressed as $\chi T$ is also shown in Figure~\ref{fig:diffuse}
and confirms the development of antiferromagnetic correlations.

\begin{table*}
\begin{centering}
\begin{tabular}{c|c|c|c|c|c|c|c|c|c|c}
\hline 
 & $J_{1}$ (K) & $J_{2}$ (K) & $J_{3a}$ (K) & $J_{3b}$ (K) & $J_{c}$ (K) & $D$ (K) & $R_{\mathrm{wp}}^{\mathrm{neutron}}$ & $R_{\mathrm{wp}}^{\mathrm{\chi T}}$ & $\mathbf{q}_{\mathrm{calc}}$ & $T_{N}^{\mathrm{calc}}$ (K)\tabularnewline
\hline 
\hline 
(a) & $-0.060(3)$ & $0^{\ast}$ & $0^{\ast}$ & $0^{\ast}$ & $-0.247(4)$ & $0.0487^{\ast}$ & $53.0$ & $2.3$ & $(0,0,\frac{3}{2})$ & $1.63$\tabularnewline
(b) & $0.009(5)$ & $-0.073(4)$ & $0^{\ast}$ & $0^{\ast}$ & $-0.196(6)$ & $0.0487^{\ast}$ & $28.0$ & $4.2$ & $(0.56,-0.56,0.56)$ & $1.36$\tabularnewline
(c) & $0.09(1)$ & $0.02(1)$ & $-0.28(1)$ & $-0.12(2)$ & $-0.06(2)$ & $0.0487^{\ast}$ & $19.2$ & $2.2$ & $(0.30,0.30,\frac{3}{2})$ & $1.57$\tabularnewline
(d) & $0.16(1)$ & $0.04(1)$ & $-0.30(1)$ & $-0.13(3)$ & $-0.08(1)$ & $0^{\ast}$ & $25.3$ & $2.0$ & $(0.28,0.28,\frac{3}{2})$ & $1.00$\tabularnewline
\hline 
\end{tabular}
\par\end{centering}
\caption{\label{tab:diffuse}Refined values of interaction parameters for different
models. Interaction parameter values are in K, and assume spins of
magnitude $\sqrt{S(S+1)}$ with $S=5/2$ for Mn$^{2+}$. Positive
values indicate ferromagnetic interactions. Parameter values held
fixed are indicated with an asterisk ($^{\ast}$).}
\end{table*}

To model these data, we consider a Hamiltonian that includes Heisenberg
exchange interactions and the long-range magnetic dipolar interaction,
\begin{equation}
H=-\sum_{i>j}J_{ij}\mathbf{S}_{i}\cdot\mathbf{S}_{j}+D\sum_{i>j}\frac{\mathbf{S}_{i}\cdot\mathbf{S}_{j}-3\left(\mathbf{S}_{i}\cdot\hat{\mathbf{r}}_{ij}\right)\left(\mathbf{S}_{j}\cdot\hat{\mathbf{r}}_{ij}\right)}{\left(r_{ij}/r_{1}\right)^{3}},\label{eq:heisenberg}
\end{equation}
where $\mathbf{S}_{i}$ is modeled as a classical vector of magnitude
$\sqrt{S(S+1)}$, $S=5/2$ is the spin quantum number of Mn$^{2+}$,
$\hat{\mathbf{r}}_{ij}=|\mathbf{r}_{j}-\mathbf{r}_{i}|/r_{ij}$ is
a unit vector parallel to the separation of spins $i$ and $j$, and
$r_{1}=3.7124(1)\,\text{\AA}$ is the nearest-neighbor distance. The
exchange interactions include the nearest-neighbor exchange $J_{1}$,
the inter-layer coupling $J_{c}$, and the further-neighbor couplings
shown in Figure~\ref{fig:crystal_structure}(c), so that $J_{ij}\in\{J_{1},J_{2},J_{3a},J_{3b},J_{c}\}$.
The magnitude of the dipolar interaction at the nearest-neighbor distance,
$D=\mu_{0}(g\mu_{\mathrm{B}})^{2}/4\pi r_{1}^{3}k_{\mathrm{B}}=0.0487$\,K,
is determined by the crystal structure. The values of the exchange
interactions were optimized against our $I(Q)$ and $\chi T$ data
(see Methods). 

We tested interaction models against our data in order of increasing
number of exchange parameters, as follows. Unless otherwise noted,
the dipolar interaction was fixed at $D=0.0487$\,K. For each exchange
model, the best fit to $I(Q)$ and $\chi T$ data is shown in Figure~\ref{fig:diffuse}.
The refined values of the exchange interactions are shown in Table~\ref{tab:diffuse},
along with the goodness of fit metric $R_{\mathrm{wp}}$, and two
quantities estimated from the Onsager-reaction-field calculation that
may indicate model quality: the predicted magnetic ordering temperature
$T_{N}^{\mathrm{calc}}$ and the predicted magnetic propagation vector
$\mathbf{q}_{\mathrm{calc}}$. 

First, we considered a minimal model in which only $J_{1}$ and $J_{c}$
were refined {[}model (a){]}. This model does not represent our neutron
data well {[}Figure~\ref{fig:diffuse}(a){]}, and the predicted propagation
vector is commensurate, in contrast to the incommensurate propagation
vector observed experimentally. Second, we refined $J_{1}$, $J_{2}$,
and $J_{c}$ parameters {[}model (b){]}. This model yields a substantially
improved fit, but some misfit is still evident in the $I(Q)$ and,
especially, the $\chi T$ data {[}Figure~\ref{fig:diffuse}(b){]}.
The calculated propagation vector is now incommensurate, but different
to the experimental one. Third, we refined $J_{1}$, $J_{2}$, $J_{3a}$,
$J_{3b}$, and $J_{c}$ parameters {[}model (c){]}. This model yields
an excellent fit to both the $I(Q)$ and $\chi T$ data {[}Figure~\ref{fig:diffuse}(c){]}.
Moreover, the calculated propagation vector, $(0.30,0.30,\frac{3}{2})$,
is close to the experimental value of $(0.3102(4),0.2646(4),\frac{3}{2})$
in the first ordered state at $0.8$\,K, and the calculated $T_{N}^{\mathrm{calc}}\approx1.6$\,K
agrees with the measured value. This refinement was stable despite
the relatively large number of free parameters; no large parameter
covariances ($\sigma_{ij}\geq80\%$) were noted, and initializing
the refinement with different parameter values yielded only one possible
local minimum, which had significantly worse $R_{\mathrm{wp}}^{\mathrm{neutron}}=23.0\%$
and $R_{\mathrm{wp}}^{\chi T}=4.1\%$. 

Our results suggest that model (c) represents well the interactions
of Na$_{2}$Mn$_{3}$Cl$_{8}$. This model has weak ferromagnetic
$J_{1}$, consistent with the Goodenough-Kanamori rules. The inter-layer
coupling $J_{c}$ is antiferromagnetic, consistent with the antiferromagnetic
layer stacking observed below $T_{N1}$. The third-neighbor couplings
$J_{3a}$ and $J_{3b}$ are antiferromagnetic and significantly larger
than $J_{1}$. Hence, Na$_{2}$Mn$_{3}$Cl$_{8}$ is an unusual system
where strong antiferromagnetic third-neighbor interactions compete
with ferromagnetic nearest-neighbor interactions. To the best of our
knowledge, a similar $J_{1}$-$J_{3}$ competition has been identified
in only one other kagome material, vesignieite \cite{Boldrin_2018}. However, this material
differs from Na$_{2}$Mn$_{3}$Cl$_{8}$ in its magnetic properties
as well as its chemistry, as it has $S=1/2$ and shows commensurate
magnetic ordering \cite{Boldrin_2018}.

Finally, we considered the relevance of the long-ranged dipolar interaction
by performing a fourth refinement in which $J_{1}$, $J_{2}$, $J_{3a}$,
$J_{3b}$, and $J_{c}$ parameters were varied, while $D$ was fixed
at zero {[}model (d){]}. This refinement yielded worse agreement with
$I(Q)$ and $\chi T$ data, and significantly underestimates the value
of $T_{N1}$ {[}Table~\ref{tab:diffuse}{]}. This result shows that
the dipolar interaction has a significant effect on the magnetic properties,
as expected since $D$ is of comparable magnitude to the exchange
interactions. However, the refined values of all parameters except
$J_{1}$ are equivalent (within $1\sigma$) for models (c) and (d),
suggesting that the effect of the dipolar term on these refinements
is largely confined to nearest neighbors.

\section{Theory and Modeling}

\subsection{Magnetic interactions from first principles}

To gain insight into the exchange interactions, we performed first-principles
calculations using density-functional theory (see Methods). The values
of the interactions calculated using DFT are given in Table~\ref{tab:dft}
for different values of the Hubbard $U$ between $0$ and $5.25$\,eV.
Based on other materials, we anticipate that $U$ is likely between
$4$ and $5.25$\,eV. 

The first-principles exchange interactions show similarities with
the experimentally-determined values, but also substantial differences.
On the one hand, the first-principles values of $J_{1}$ and $J_{c}$
are ferromagnetic and antiferromagnetic, respectively, consistent
with the values fitted to experimental data. The magnitudes of $J_{1}$
and $J_{c}$ for $U=5.25$\,eV are also comparable to the experimentally-determined
magnitudes, in contrast to a previous DFT study that reported interactions
larger than $30$\,K . On the other hand, the first-principles values
of $J_{2}$, $J_{3a}$, and $J_{3b}$ are \emph{opposite }to the experimentally-determined
values; moreover, the calculated magnitudes of these interactions
are very large compared to the other interactions.

We carefully checked whether the first-principles results could be
consistent with our experimental data. Taking $U=5.25$\,eV, we
calculate the Weiss temperature as $\theta_{\mathrm{DFT}}=\frac{4}{3}S(S+1)[J_{1}+J_{2}+J_{c}+J_{3a}+J_{3b}/2]=3.0$\,K.
Hence, DFT predicts a ferromagnetic Weiss temperature, which is not
consistent with the antiferromagnetic value ($\theta=-4.6(1)$\,K)
measured experimentally. We also estimate the magnetic ordering temperature
to be $4.8$\,K, which is much larger than the experimental value
of $1.6$\,K. Finally, we performed additional refinements to neutron
and $\chi T$ data as described in Section~\ref{subsec:diffuse},
except we constrained the signs of the exchange interactions to be
the same as those from DFT, while allowing their magnitudes to refine
freely. These refinements yielded $J_{3a}=J_{3b}\approx0$, essentially
reproducing the results of model (b) in Section~\ref{subsec:diffuse}. 

We therefore conclude that the DFT results are not fully consistent
with our experimental data, making Na$_{2}$Mn$_{3}$Cl$_{8}$ a model
material for benchmarking developments in first-principles calculations.
The reason for the inaccuracy of the DFT exchange interactions beyond
nearest-neighbors is not yet clear; an interesting possibility is
that it may relate to the neglect of the Stoner coupling on the Cl
ligand sites, as recently proposed in the related material NaMnCl$_{3}$
\citep{Solovyev_2022}.

\begin{table}
\begin{centering}
\begin{tabular}{c|c|c|c|c|c}
\hline 
$U$ (eV) & $J_{1}$ (K) & $J_{2}$ (K) & $J_{3a}$ (K) & $J_{3b}$ (K) & $J_{c}$ (K)\tabularnewline
\hline 
\hline 
$0$ & $0.516$ & $-1.321$ & $0.801$ & $1.045$ & $-0.040$\tabularnewline
$2.00$ & $0.288$ & $-0.745$ & $0.525$ & $0.591$ & $-0.029$\tabularnewline
$4.00$ & $0.194$ & $-0.507$ & $0.411$ & $0.406$ & $-0.024$\tabularnewline
$5.25$ & $0.163$ & $-0.430$ & $0.377$ & $0.346$ & $-0.023$\tabularnewline
\hline 
\end{tabular}
\par\end{centering}
\caption{\label{tab:dft}Values of interaction parameters obtained from density-functional
theory simulations for different values of the Hubbard $U$.}
\end{table}

\subsection{Origin of incommensurate ordering\label{subsec:incommensurate}}

\begin{figure*}
\includegraphics{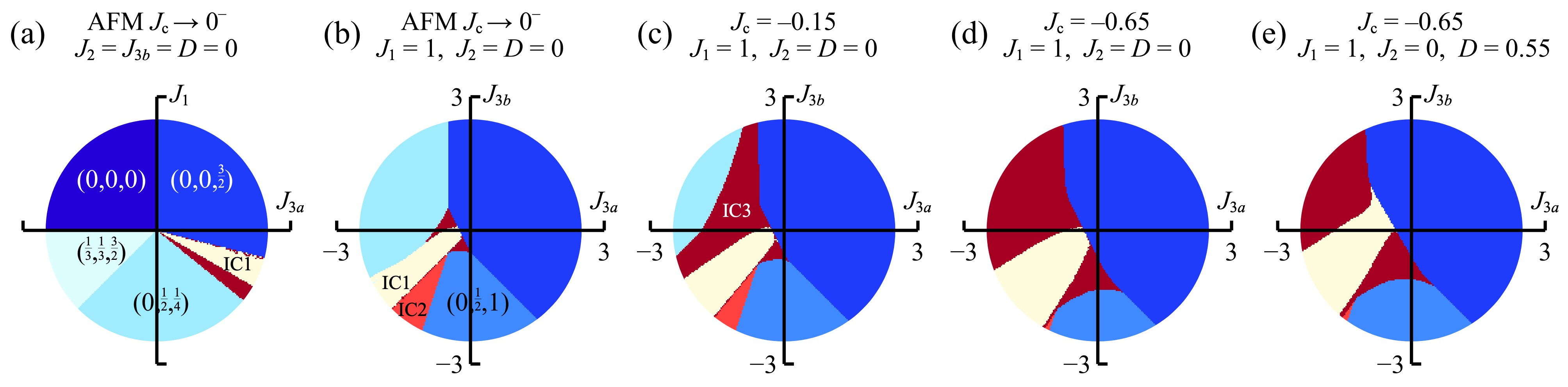}

\caption{\label{fig:phase_diagram}Mean-field phase diagrams for the kagome
lattice of Na$_{2}$Mn$_{3}$Cl$_{8}$. The values of fixed interaction
parameters are given above each phase diagram (a)--(e), and axes
are labeled with the variable interaction parameters. Different magnetic
propagation vectors $\mathbf{q}$ are indicated by different colors,
with the \textbf{q} corresponding to each color labeled in the leftmost
phase diagram in which it occupies a wide phase space. The propagation
vectors include $(0,0,0)$, $(0,0,\frac{3}{2})$, $(\frac{1}{3},\frac{1}{3},\frac{3}{2})$,
$(0,\frac{1}{2},\frac{1}{4})$, $(0,\frac{1}{2},1)$, and three incommensurate
vectors, $\mathrm{IC1}=(q,q,\frac{3}{2})$, $\mathrm{IC2}=(q,q,0)$,
and $\mathrm{IC2}=(0,q,r)$. Note that the interlayer coupling $J_{c}$
is antiferromagnetic for all phase diagrams, and the nearest-neighbor
coupling $J_{1}$ is ferromagnetic for (b)--(e).}
\end{figure*}

In this section, we discuss the origin of the multiple incommensurate
ordering transitions in Na$_{2}$Mn$_{3}$Cl$_{8}$, using a combination
of field-theoretic and Monte Carlo simulations.

Incommensurate magnetic structures are relatively uncommon in kagome
antiferromagnets. For example, to the best of our knowledge, all known
jarosite minerals that exhibit long-range order have either $(0,0,0)$
or $(0,0,\frac{3}{2})$ propagation vectors (see \citep{Mendels_2011}
and references therein). Similarly, commensurate states are observed
for many other insulating materials in which the kagome lattice is
undistorted or slightly distorted; for example, MgFe$_{3}$(OH)$_{6}$Cl$_{2}$
with $\mathbf{q}=(0,0,\frac{3}{2})$ \citep{Fujihala_2017}, centennialite
CaCu$_{3}$(OH)$_{6}$Cl$_{2}\cdot$0.6H$_{2}$O \citep{Iida_2020}, CdCu$_{3}$(OH)$_{6}$(NO$_{3}$)$_{2}\cdot$0.6H$_{2}$O
\citep{Ihara_2022}, Nd$_{3}$Sb$_{3}$Mg$_{2}$O$_{14}$ \citep{Scheie_2016},
and Sr-vesignieite SrCu$_{3}$V$_{2}$O$_{8}$(OH)$_{2}$ with $\mathbf{q}=(0,0,0)$, $\alpha$-Cu$_{3}$Mg(OH)$_{6}$Br$_{2}$ \citep{Wei_2019} and
YCu$_{3}$(OH)$_{6}$Cl$_{3}$ with $\mathbf{q}=(0,0,\frac{1}{2})$
\citep{Zorko_2019}, and Ba-vesignieite BaCu$_{3}$V$_{2}$O$_{8}$(OH)$_{2}$
with $\mathbf{q}=(\frac{1}{2},0,0)$ \citep{Boldrin_2018,Okamoto_2009}.
By contrast, the distorted-kagome material
Ba$_{2}$Mn$_{3}$F$_{11}$ is one of the only insulating kagome materials
with incommensurate magnetic ordering \citep{Hayashida_2018}. Incommensurate
modulations are more frequently observed in metallic kagome systems,
such as Tb$_{3}$Ru$_{4}$Al$_{12}$ \citep{Rayaprol_2019} and YMn$_{6}$Sn$_{6}$,
the latter of which undergoes an incommensurate-to-commensurate transition
on cooling \citep{Ghimire_2020}.

To understand the preference for kagome magnets to form commensurate
structures, and the conditions where incommensurate structures may
appear, we use the reciprocal-space mean-field approximation introduced
in Section~\ref{subsec:methods_diffuse_field} to investigate the
stability of different phases as a function of the interactions $J_{1}$,
$J_{3a}$, $J_{3b}$, and $J_{c}$ {[}Figure~\ref{fig:crystal_structure}(c){]}.
Throughout large regions of this interaction space, the classical
ground state is one of the commensurate ``regular magnetic orders''
described in Ref.~\citep{Messio_2011}. Of the models previously
investigated theoretically, the most relevant one to Na$_{2}$Mn$_{3}$Cl$_{8}$
is the $J_{1}$-$J_{3a}$ Heisenberg model studied in Refs.~\citep{Grison_2020,Li_2022}.
The phase diagram for this model is shown in Figure~\ref{fig:phase_diagram}(a),
and contains five phases: ferromagnetic layers with antiferromagnetic
stacking {[}$\mathbf{q}=(0,0,\frac{3}{2})${]}, $\mathbf{q=0}$ antiferromagnet,
$\sqrt{3}\times\sqrt{3}$ antiferromagnet {[}$\mathbf{q}=(\frac{1}{3},\frac{1}{3},\frac{3}{2})${]},
three-sublattice antiferromagnet {[}$\mathbf{q}=(0,\frac{1}{2},\frac{1}{4})${]},
and an incommensurate region. This result reproduces the result of
Ref.~\citep{Grison_2020} for isolated kagome planes, except that
we include a small antiferromagnetic inter-layer coupling $J_{c}\rightarrow0^{-}$
to stabilize three-dimensional ordering. 

While the $J_{1}$-$J_{3a}$ phase diagram is relatively complicated,
it is nevertheless simpler than our model for Na$_{2}$Mn$_{3}$Cl$_{8}$,
which also includes significant $J_{c}$, $J_{3b}$, and dipolar couplings.
We therefore extended the $J_{1}$-$J_{3a}$ phase diagram to consider
the effects of these additional couplings, which are needed for a
full description of our Na$_{2}$Mn$_{3}$Cl$_{8}$ data. Notably,
for all models, antiferromagnetic $J_{3a}$ is necessary to stabilize
incommensurate ordering with $\mathbf{q}=(q,q,\frac{3}{2})$. In Figure~\ref{fig:phase_diagram}(b),
we fix ferromagnetic $J_{1}=1$ and consider the phase diagram in
the $J_{3a}$-$J_{3b}$ plane for antiferromagnetic $J_{c}\rightarrow0^{-}$.
Nonzero $J_{3b}$ has a dramatic effect on the phase diagram; in particular,
including antiferromagnetic $J_{3b}$ extends the stability region
of the incommensurate phases observed for antiferromagnetic $J_{3a}$.
Figure~\ref{fig:phase_diagram}(b)--(d) show the effect of increasing
the magnitude of $J_{c}$, the antiferromagnetic interlayer coupling
($J_{c}\rightarrow0^{-}$, $-0.15$, and $-0.65$, respectively, in
the same units as $J_{1}$). The effect of increasing $|J_{c}|$ is
to increase further the region of phase space in which incommensurate
order is stable within the mean-field approximation. Finally, in Figure~\ref{fig:phase_diagram}(e),
we show the $J_{3a}$-$J_{3b}$ phase diagram including the long-range
dipolar interaction $D/J_{1}\approx0.55$ appropriate for Na$_{2}$Mn$_{3}$Cl$_{8}$.
The inclusion of $D$ has a relatively small effect on the positions
of the phase boundaries. 

\begin{figure}
\begin{centering}
\includegraphics{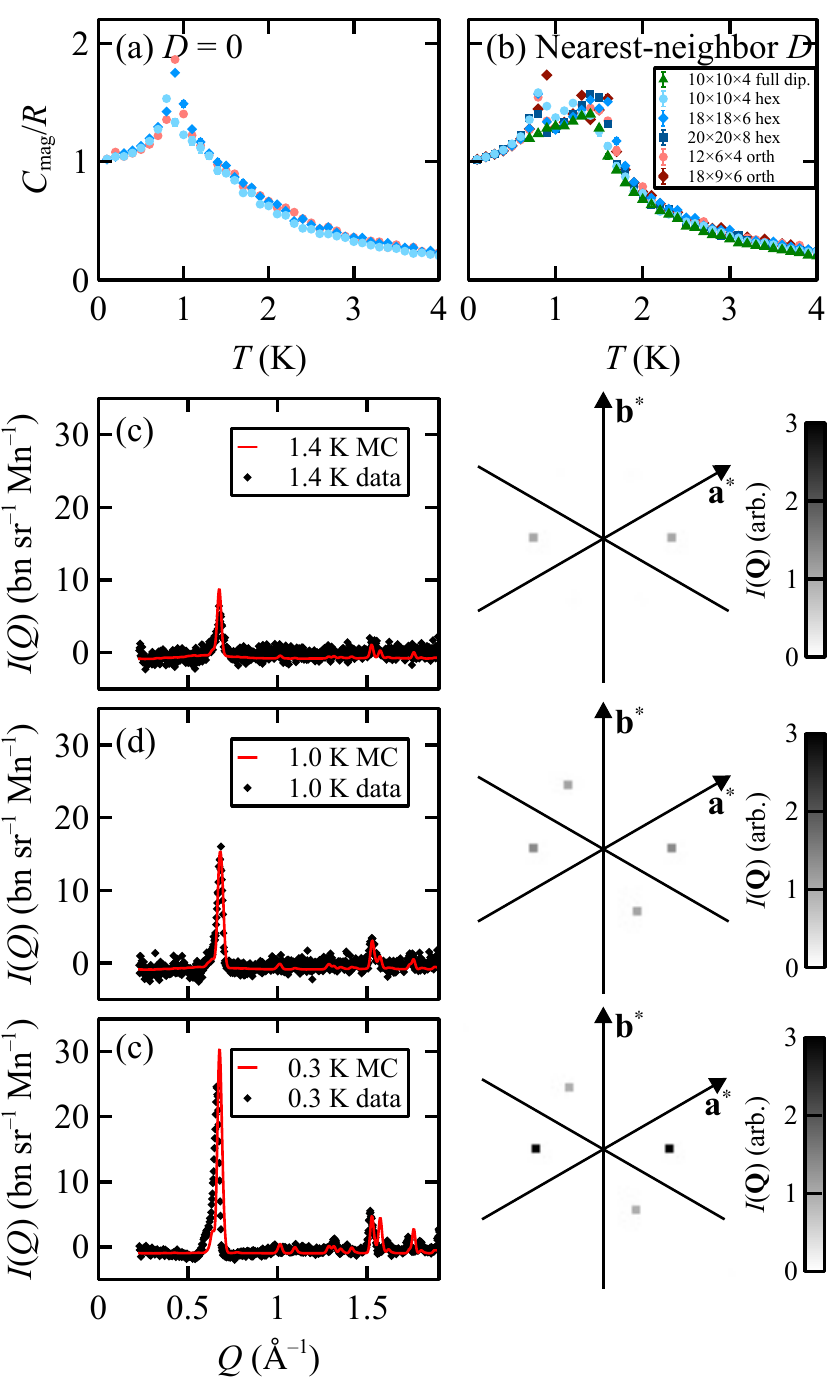}
\par\end{centering}
\centering{}\caption{\label{fig:mc}(a) Magnetic specific heat calculated from Monte Carlo
simulations of the exchange parameters of model (c) in Table~\ref{tab:diffuse},
taking the dipolar interaction $D=0$. (b) As (a), except the dipolar
interaction is included, either for all neighbors (green triangles) or for nearest neighbors only (all other points). The simulated
system sizes are as shown in the key in (b). (c) Magnetic diffraction
patterns at $T=1.4$\,K, showing calculated powder diffraction profile
from Monte Carlo simulation (red line, left), experimental powder-diffraction
data (black circles, left), and calculated single-crystal diffraction
pattern from Monte Carlo simulation (grayscale plot, right). (d) As
(c), except at $T=1.0$\,K. (e) As (c), except at $T=0.3$\,K.}
\end{figure}

The reciprocal-space mean-field theory provides a useful overview
of the phase space, but has several important limitations. First,
for a non-Bravais lattice such as kagome, it only determines a lower
bound on the energy of the ground state. As discussed in Ref.~\citep{Grison_2020},
for the incommensurate region of the $J_{1}$-$J_{3a}$ phase diagram,
a physical spin configuration could not be identified that reached
this lower bound; hence, the actual magnetic ground state is uncertain
in this region. Second, since this theory considers instabilities
of the paramagnetic phase, it predicts only the propagation vector
of the first ordered state that develops on cooling; it provides no
information about the possibility of multiple phase transitions, as
are observed experimentally in Na$_{2}$Mn$_{3}$Cl$_{8}$. 

We performed classical Monte Carlo simulations to address these limitations.
Since the periodicity of an incommensurate magnetic structure does
not ``fit'' within any finite-sized configuration, finite-size artifacts
are encountered, which can be reduced by studying relatively large
system sizes. However, the long-ranged nature of the magnetic dipolar
interaction makes large system sizes computationally expensive. We
therefore consider first an approximation to the full Hamiltonian,
Eq.~(\ref{eq:heisenberg}), where we simulate the parameters that
best describe our diffuse-scattering data {[}model (c) in Table~\ref{tab:diffuse}{]},
but truncate the dipolar interaction $D$ at the nearest-neighbor
distance; we will call this the ``nearest-neighbor dipolar model''.
For comparison, we also simulated the same model (c) except with $D=0$.
To identify finite-size effects, we considered different system dimensions
from $10\times10\times4$ hexagonal unit cells ($3600$ spins) to
$20\times20\times8$ hexagonal unit cells ($28800$ spins). For the
$10\times10\times4$ and $20\times20\times8$ simulations only, we
slightly adjusted the model (c) interaction parameters to stabilize
$\mathbf{q}=(\frac{3}{10},\frac{3}{10},\frac{3}{2})$ ordering, which
is commensurate with the system size; this was achieved by multiplying
the best-fit values of $J_{3a}$ and $J_{3b}$ by $0.936$. To investigate
the effect of a different system geometry, we defined a orthogonal
unit cell with axes $\mathbf{a}_{\mathrm{o}}=\mathbf{a}$, $\mathbf{b}_{\mathrm{o}}=\mathbf{a}+2\mathbf{b}$,
and $\mathbf{c}_{\mathrm{o}}=\mathbf{c}$, and performed simulations
of $12\times6\times4$ and $18\times9\times6$ orthogonal unit cells
($5184$ and $17496$ spins, respectively). Simulations were run for
up to $4.1\times10^{6}$ moves per spin at low temperatures, where
a single move involved one microcanonical (over-relaxation) update
followed by a proposed spin rotation of a randomly-chosen spin, which
was accepted or rejected according to the Metropolis criterion. Measurements
of the autocorrelation function showed that these conditions allowed
the system to decorrelate at all temperatures above $0.1$\,K. Simulations
including the long-ranged dipolar interaction, implemented using Ewald
summation \citep{Wang_2001}, were also performed for a small system
size of $10\times10\times4$ hexagonal unit cells, without over-relaxation
updates.

Results of our Monte Carlo simulations are shown in Figure~\ref{fig:mc}.
For the model with $D=0$, a sharp anomaly indicating a single magnetic
phase transition is observed at $\approx0.9$\,K; we do not consider
the low-temperature state here. The nearest-neighbor dipolar model
shows a more complex temperature evolution. In all our simulations,
sharp specific heat anomaly is observed at $\approx0.9$\,K, with
a second feature between $1.3$ and $1.6$\,K that is resolved as
either a single broadened peak or two peaks close in temperature,
depending on system dimensions. Hence, unlike the Heisenberg model,
the nearest-neighbor dipolar model shows at least two magnetic phase
transitions, in qualitative agreement with the experimental data for
Na$_{2}$Mn$_{3}$Cl$_{8}$. Properties of the magnetic phases obtained
for a model of $18\times9\times6$ orthogonal unit cells are shown
at $1.4$, $1.0$, and $0.3$\,K, in Figure~\ref{fig:mc}(c), (d),
and (e) respectively. The phases observed at $1.0$ and $0.3$\,K
are resolved for all other system sizes and geometries. However, the
$1.4$\,K phase is not resolved in the $20\times20\times8$ simulation,
suggesting its appearance for some other system sizes may be a finite-size
artifact. The calculated magnetic powder diffraction patterns show
remarkably good agreement with our experimental powder-diffraction
data, especially at $1.4$ and $1.0$\,K {[}Figure~\ref{fig:mc}(c)--(e){]}.
Calculations of the single-crystal magnetic diffraction patterns reveal
magnetic Bragg peaks corresponding to a single incommensurate wavevector
at $1.4$\,K, indicating a single-$\mathbf{q}$ magnetic structure
at this temperature {[}Figure~\ref{fig:mc}(c){]}. Remarkably, however,
the same calculation shows magnetic Bragg peaks corresponding to two
wavevectors at $1.0$ and $0.3$\,K. The intensity of each wavevector
is approximately equal at $1.0$\ K but significantly different at
$0.3$\,K {[}Figure~\ref{fig:mc}(d) and (e){]}. The same effect
was observed across all our simulations at $1.0$ and $0.3$\,K,
suggesting this is likely not an artifact due to domain formation,
but instead indicates the formation of double-$\mathbf{q}$ states
in the Monte Carlo simulations. Our simulations of the long-ranged
dipolar model also suggest a possible change in magnetic structure
below approximately $1.0$\,K, although a second transition is not
clearly resolved in the heat capacity for this small simulation size
{[}Figure~\ref{fig:mc}(b){]}. For this model, the magnetic structure
is clearly 2-$\mathbf{q}$ only below $1.0$\,K.

Our results suggest the enticing possibility that the ordered incommensurate
states may, in fact, be multi-$\mathbf{q}$ structures rather than
single-$\mathbf{q}$ helices. Given the good agreement of our microscopic
model with powder-diffraction data and its correct prediction of multiple
phase transitions, this scenario is certainly possible. Further theoretical
studies including the long-ranged dipolar interaction would be useful
to elucidate the relative stabilities of single-$\mathbf{q}$ and
multi-$\mathbf{q}$ states, which may be close in energy.

\section{Conclusions}

Our neutron-diffraction study reveals that Na$_{2}$Mn$_{3}$Cl$_{8}$
shows novel magnetic behavior. Unusually for a kagome antiferromagnet,
it shows incommensurate ordering; even more unusually, it exhibits
multiple incommensurate magnetic phases, which form at $1.6$ and
$0.6$\,K. To the best of our knowledge, ordering wavevectors of
the form $(q_{x},q_{y},\frac{3}{2})$, as observed in Na$_{2}$Mn$_{3}$Cl$_{8}$,
have not previously been observed in insulating kagome magnets. As
such, Na$_{2}$Mn$_{3}$Cl$_{8}$ significantly expands the known
range of magnetic behavior on the kagome lattice. 

We investigated the magnetic interactions that drive incommensurate
ordering in Na$_{2}$Mn$_{3}$Cl$_{8}$ using experiment-driven and
first-principles approaches. By fitting the magnetic diffuse scattering
measured above the magnetic ordering temperature, we showed that the
magnetic interactions extend to third-nearest neighbors. Antiferromagnetic
third-neighbor interactions $J_{3a}$ and $J_{3b}$ are the largest
terms in the Hamiltonian, and compete with ferromagnetic nearest-neighbor
interactions $J_{1}$. Using a mean-field theory, we showed that antiferromagnetic
$J_{3a}$, $J_{3b}$, and interlayer couplings extends the stability
region of incommensurate ordering in a model with ferromagnetic $J_{1}$.
Our experimentally-determined interactions could not be fully reproduced
by DFT calculations, which predict ferromagnetic $J_{3a}$ and $J_{3b}$,
inconsistent with our experimental data. This material may therefore
be a useful test case for advancements in first-principles methodologies. 

Using magnetic Rietveld refinement, we showed that the magnetic Bragg
profiles of the two incommensurate magnetic phases are well described
by single-$\mathbf{q}$ helical structures. These are cycloidal helices,
in which the spins and the propagation vector $\mathbf{q}$ both have
a component in the $ab$-plane. Due to the limitations of powder data,
however, other structures can give equivalent or slightly better agreement
with the experimental pattern. We showed that single-$\mathbf{q}$
sine structures are highly unlikely at $0.8$ and $0.3$\,K, since
some sites would have unphysically large magnitudes of the ordered
magnetic moment. However, we were not able to rule out multi-$\mathbf{q}$
structures, which are generally indistinguishable from their single-$\mathbf{q}$
analogs in powder diffraction measurements. This issue is especially
relevant here, because Monte Carlo simulations of our experimentally-determined
interaction model show multiple magnetic phases transitions, in qualitative
agreement with the experimental data, and indicate that two of the
phases obtained are 2-$\mathbf{q}$ states. Further experiments
would therefore be valuable to distinguish between single-$\mathbf{q}$
and double-$\mathbf{q}$ states. These experiments could include single-crystal
neutron diffraction under applied magnetic field, or inelastic neutron
scattering. The growth of large single crystals of Na$_{2}$Mn$_{3}$Cl$_{8}$
would facilitate such measurements and potentially shed further light
on the nature of the spin texture in Na$_{2}$Mn$_{3}$Cl$_{8}$.
\begin{acknowledgments}
We are grateful to Andrew Christianson (Oak Ridge National Laboratory)
for valuable discussions. This work was supported by the U.S. Department
of Energy (DOE), Office of Science, Basic Energy Sciences, Materials
Sciences and Engineering Division, and used resources at the High
Flux Isotope Reactor, a DOE Office of Science User Facility operated
by the Oak Ridge National Laboratory. 
\end{acknowledgments}


\end{document}